\documentclass[12pt]{article}

\pdfoutput=1

\usepackage{color}

\usepackage{amsmath,amssymb,latexsym}
\usepackage[utf8]{inputenc}  

\usepackage{cite}

\usepackage[T1]{fontenc}
\usepackage{times}

\usepackage{epstopdf}

\usepackage{graphics,epsfig}

\usepackage{caption}
\usepackage{fdsymbol}

\usepackage{indentfirst}

\usepackage{colortbl}

\usepackage{xcolor}

\usepackage{enumitem}

\usepackage{circledsteps}

\definecolor{olive}{rgb}{0.3, 0.4, .1}
\definecolor{fore}{RGB}{249,242,215}
\definecolor{back}{RGB}{51,51,51}
\definecolor{title}{RGB}{255,0,90}
\definecolor{dgreen}{rgb}{0.,0.6,0.}
\definecolor{gold}{rgb}{1.,0.84,0.}
\definecolor{JungleGreen}{cmyk}{0.99,0,0.52,0}
\definecolor{BlueGreen}{cmyk}{0.85,0,0.33,0}
\definecolor{RawSienna}{cmyk}{0,0.72,1,0.45}
\definecolor{Magenta}{cmyk}{0,1,0,0}
\definecolor{darkred}{rgb}{0.55,0.,0.}
\definecolor{coxgreen}{rgb}{0.4,0.5,0.08}
\definecolor{dolgreen}{rgb}{0.33,0.42,0.18}

\definecolor{greym}{RGB}{76,76,76}

\definecolor{UniBlue}{RGB}{83,91,170}

\let\a=\alpha \let\b=\beta \let\g=\gamma \let\d=\delta \let\e=\epsilon
\let\i=\iota \let\k=\kappa
\let\l=\lambda \let\m=\mu \let\n=\nu \let\x=\xi \let\p=\pi 
\let\s=\sigma 
 \let\f=\phi  \let\y=\psi
 
        \let\Th=\Theta 
\let\X=\Xi  \let\S=\Sigma  \let\Y=\Psi
 
\let\la=\label  
  
 \def\bd{\begin{document}} \def\ed{\end{document}}
\def\ds{\documentstyle} \let\fr=\frac \let\bl=\bigl \let\br=\bigr
\let\Br=\Bigr \let\Bl=\Bigl
\let\bm=\bibitem
\let\na=\nabla
\def\tU{{\widetilde U}}
\let\pa=\partial \let\ov=\overline
\def\ie{{\it i.e.\ }}
\newcommand{\be}{\begin{equation}}
\newcommand{\ee}{\end{equation}}
\def\ba{\begin{array}}
\def\ea{\end{array}}

\def\bei{\begin{itemize}}
\def\eei{\end{itemize}}

\def\ben{\begin{enumerate}}
\def\een{\end{enumerate}}

\def\ft#1#2{{\textstyle{{\scriptstyle #1}\over {\scriptstyle #2}}}}
\def\fft#1#2{{#1 \over #2}}
\def\F#1#2{{ F_{#1}^{(#2)} }}
\def\cF#1#2{{ {\cal F}_{#1}^{(#2)} }}

\def\R{{\bf R}}
\def\sst#1{{\scriptscriptstyle #1}}
\def\oneone{\rlap 1\mkern4mu{\rm l}}
\def\e7{E_{7(+7)}}
\def\td{\tilde}
\def\wtd{\widetilde}
\def\im{{\rm i}}
\def\bog{Bogomol'nyi\ }
\newcommand{\ho}[1]{$\, ^{#1}$}
\newcommand{\hoch}[1]{$\, ^{#1}$}
\newcommand{\bea}{\begin{eqnarray}}
\newcommand{\eea}{\end{eqnarray}}
\newcommand{\ra}{\rightarrow}
\newcommand{\lra}{\longrightarrow}
\newcommand{\Lra}{\Leftrightarrow}
\newcommand{\ap}{\alpha^\prime}
\newcommand{\bp}{\tilde \beta^\prime}
\newcommand{\cB}{{\cal B}}
\newcommand{\cO}{{\cal O}}
\newcommand{\vecx}{\vec{x}}
\newcommand{\vecy}{\vec{y}}
\newcommand{\vecp}{\vec{p}}
\newcommand{\vecq}{\vec{q}}
\newcommand{\tr}{{\rm tr} }
\newcommand{\Tr}{{\rm Tr} }
\newcommand{\NP}{Nucl. Phys. }

\newcommand{\cL}{{\cal L}}
\newcommand{\cA}{{\cal A}}
\newcommand{\cT}{{\cal T}}
\newcommand{\cD}{{\cal D}}
\newcommand{\cH}{{\cal H}}

\def\th{\theta}

\def\sst#1{{\scriptscriptstyle #1}}
\def\0{{\sst{(0)}}}
\def\1{{\sst{(1)}}}
\def\2{{\sst{(2)}}}
\def\3{{\sst{(3)}}}
\def\4{{\sst{(4)}}}
\def\5{{\sst{(5)}}}
\def\6{{\sst{(6)}}}
\def\7{{\sst{(7)}}}
\def\8{{\sst{(8)}}}
\def\9{{\sst{(9)}}}
\def\p{{\sst{(p)}}}
\def\q{{\sst{(q)}}}

\def\ssa{{\sst{(\alpha)}}}
\def\ssb{{\sst{(\beta)}}}
\def\ssg{{\sst{(\gamma)}}}
\def\j{{\sst{(j)}}}

\def\ve{\varepsilon}
\def\vf{\varphi}
\def\F{\Phi}
\def\wg{\wedge}

\def\thb{\bar{\theta}}
\def\Thb{\bar{\Theta}}
\def\barp{\bar{p}}
\def\barq{\bar{q}}
\def\barc{\bar{c}}
\def\bard{\bar{d}}
\def\e{\epsilon}

\def \bi{\bibitem}
\def \la {\label}

\def \l {\lambda}
\def\foot{\footnote}
\def \tl  {{\tilde \l}}
\def \sql {{\sqrt \l}}
\def \adss {$AdS_5 \times S^5$\ }
\newcommand{\rf}[1]{(\ref{#1})}
\def \ov {\over}

\def\Th{\Theta}
\def\vth{\vartheta}
\def\btheta{{\bar\theta}}
\def\ttheta{{{\tilde\theta}}}
\def\bttheta{{{\bar\ttheta}}}
\def\vth{\vartheta}

\def\ra{\rightarrow}
\def\N{{\cal N}}
\def\uM{\underline{M}}
\def\uA{\underline{A}}
\def\uN{\underline{N}}
\def\uP{\underline{P}}
\def\ua{\underline{a}}
\def\ub{\underline{b}}
\def\uc{\underline{c}}
\def\ud{\underline{d}}
\def\ue{\underline{e}}
\def\uf{\underline{f}}
\def\ui{\underline{i}}
\def\uj{\underline{j}}
\def\uk{\underline{k}}
\def\ual{\underline{\alpha}}
\def\ube{\underline{\beta}}
\def\um{\underline{m}}
\def\un{\underline{n}}
\def\up{\underline{p}}
\def\uq{\underline{q}}
\def\ur{\underline{r}}
\def\us{\underline{s}}

\def\umu{\underline{\mu}}
\def\unu{\underline{\nu}}
\def\ula{\underline{\l}}
\def\uka{\underline{\k}}
\def\usi{\underline{\s}}
\def\urh{\underline{\r}}
\def\cc{\circ}
\def\eqv{\equiv}

\def\ni{\noindent}

\def\Ep{E^{{}^{(+)}}}
\def\Em{E^{{}^{(-)}}}

\def\Mp{M^{{}^{(+)}}}
\def\Mm{M^{{}^{(-)}}}

\def \ha{{1\ov 2}}

\def\r{\rho}

\def\Y{{\rm Y}}
\def\X{{\rm X}}
\def\tY{\tilde{\rm Y}}
\def\tX{\tilde{\rm X}}
\def\dY{\dot{\rm Y}}
\def\dX{\dot{\rm X}}

\def \J {\mathcal{J}}
\def \del {\partial}

\def\dF{\dot{F}}
\def\dG{\dot{G}}
\def\df{\dot{f}}
\def\dx{\dot{x}}
\def \E {{\cal E}}
\def \S {{\cal S}}
\def \J {{\cal J}}

\def\ms{\mathcal{S}}
\def\mj{\mathcal{J}}
\def\soj{\fr{\ms}{\mj}}
\def \R {{\bf R}}
\def \om {\omega}
\def \bE {\bar E}
\def \x {{\cal X}}

\def \bi{\bibitem}
\def \la {\label}

\def \l {\lambda}
\def\foot{\footnote}
\def \tl  {{\tilde \l}}
\def \sql {{\sqrt \l}}
\def \adss {$AdS_5 \times S^5$\ }
\def \ov {\over}

\def \varpi {{\rm w}}

\def\thb{\bar{\theta}}
\def\Thb{\bar{\Theta}}
\def\mb{\bar{\m}}
\def\ab{\bar{\a}}
\def\zb{\bar{z}}
\def\psib{\bar{\psi}}
\def\barp{\bar{p}}
\def\barq{\bar{q}}
\def\barc{\bar{c}}
\def\bard{\bar{d}}
\def\e{\epsilon}
\def\wb{\bar{w}}
\def\lb{\bar{\l}}
\def\Jb{\bar{J}}
\def\Nb{\bar{N}}
\def\Zb{\bar{Z}}
\def\pab{\bar{\pa}}

\def\At{\tilde{A}}
\def\Bt{\tilde{B}}
\def\Ct{\tilde{C}}
\def\Dt{\tilde{D}}
\def\Et{\tilde{E}}
\def\Ft{\tilde{F}}
\def\Gt{\tilde{G}}
\def\Ht{\tilde{H}}
\def\Mt{\tilde{M}}
\def\Rt{\tilde{R}}
\def\at{\tilde{a}}
\def\bt{\tilde{b}}
\def\ct{\tilde{c}}
\def\dt{\tilde{d}}
\def\et{\tilde{e}}
\def\ft{\tilde{f}}
\def\gt{\tilde{g}}
\def\mt{\tilde{\mu}}
\def\nt{\tilde{\nu}}

\def\asth{\hat{*}}
\def\phh{\hat{\phi}}

\def\bA{{\bf A}}

\def\ola{\overleftarrow}
\def\ora{\overrightarrow}
\def\alt{\tilde{\a}}

\def\eh{\hat{e}}
\def\eph{\hat{\e}}
\def\ph{\hat{p}}
\def\alh{\hat{\a}}
\def\beh{\hat{\b}}
\def\gah{\hat{\g}}
\def\Fh{\hat{F}}
\def\muh{\hat{\m}}
\def\nuh{\hat{\n}}
\def\thh{\hat{\th}}
\def\dh{\hat{d}}
\def\ih{\hat{i}}
\def\jh{\hat{j}}
\def\kh{\hat{k}}
\def\deh{\hat{\d}}
\def\wh{\hat{w}}
\def\lah{\hat{\l}}
\def\Ah{\hat{A}}
\def\Ch{\hat{C}}
\def\Omh{\hat{\Omega}}

\def\xh{\hat{x}}

\def\ps{\rlap{\, /}\;\,p }
\def\ks{\rlap{\, /}\;\,k }

\def\gym{g_{YM}}

\def\adot{\dot{a}}
\def\bdot{\dot{b}}
\def\bpa{\bar{\pa}}

\def\pr{\prime}
\def\ssk{\medskip}
\def\bsk{\bigskip}

\def\clb{\color{blue}}
\def\clr{\color{red}}
\def\clv{\color{violet}}
\def\clg{\color{dolgreen}}
\def\clu{\color{UniBlue}}
\def\cly{\color{yellow}}

\def\t{\tau}
\def\cM{\mathcal{M}}
\def\S{\Sigma}

\def\N{\nabla}

\def\cR{\mathcal{R}}
\def\cL{\mathcal{L}}

\def\hb{\hbar}
\def\an{\hat{a}}
\def\ac{\hat{a}^\dag}
\def\hp{\hat{p}}

\def\Ec{{\cal E}}

\def\hP{{\hat \Psi}}
\def\hE{{\hat E}}
\def\hC{{\hat C}}

\DeclareFontFamily{U}{FdSymbolA}{}
\DeclareFontShape{U}{FdSymbolA}{m}{n}{
    <-> s * [1] FdSymbolA-Book
}{}
\DeclareFontShape{U}{FdSymbolA}{m}{b}{
    <-> s * [1] FdSymbolA-Medium
}{}
\DeclareSymbolFont{fdsymbols}{U}{FdSymbolA}{m}{n}
\SetSymbolFont{fdsymbols}{bold}{U}{FdSymbolA}{m}{b}

\DeclareMathSymbol{\medtriangleright}{\mathbin}{fdsymbols}{86}
\DeclareMathSymbol{\medtriangleup}{\mathbin}{fdsymbols}{87}
\DeclareMathSymbol{\medtriangleleft}{\mathbin}{fdsymbols}{88}
\DeclareMathSymbol{\nabla}{\mathbin}{fdsymbols}{89}

\begin{document}

\title{
{\Large{\bf 
Non-stationary SQM/IST Correspondence and ${\cal CPT}/{\cal PT}$-invariant paired Hamiltonians on the line
}}
}
{
\author{V.P. Berezovoj$\,^{\spadesuit}$\footnote{berezovoj@kipt.kharkov.ua}
\,\, \,and  \,A.J. Nurmagambetov$\,^{\spadesuit,\vardiamondsuit,\varheartsuit}$\footnote{ajn@kipt.kharkov.ua; a.j.nurmagambetov@gmail.com}
\\ \\
$\,^{\spadesuit}${ \it {\normalsize Akhiezer Institute for Theoretical Physics of NSC KIPT}}
\\
{ \it {\normalsize 1 Akademichna St., Kharkiv 61108, Ukraine} }
\\
$\,^{\vardiamondsuit}${ \it {\normalsize Department of Physics \& Technology, Karazin Kharkiv National University,}}
\\
{ \it {\normalsize 4 Svobody Sq., Kharkiv 61022, Ukraine} }
\\
$\,^{\varheartsuit}${ \it {\normalsize Usikov Institute of Radiophysics and Electronics}}\\
{ \it {\normalsize 12 Ak. Proskury, Kharkiv 61085, Ukraine} }
}

\date{}

\maketitle

\abstract{
We fill some of existed gaps in the correspondence between Supersymmetric Quantum Mechanics and the Inverse Scattering Transform by extending the consideration to the case of paired stationary and non-stationary Hamiltonians. We formulate the corresponding to the case Goursat problem and explicitly construct the kernel of the non-local Inverse Scattering Transform, which solves it.  As a result, we find the way of constructing non-hermitian Hamiltonians from the initially hermitian ones, that leads, in the case of real-valued spectra of both potentials, to pairing of ${\cal CPT/PT}$-invariant Hamiltonians. The relevance of our proposal to Quantum Optics and optical waveguides technology, as well as to non-linear dynamics and Black Hole Physics is briefly discussed.

}

\bsk\bsk
PACS: 03.65.-w, 12.60.Jv, 02.30.lk, 02.30.zz  

\ssk
\noindent {\it Keywords:} Supersymmetric Quantum Mechanics, Inverse Scattering Transform, ${\cal PT}$ symmetric Quantum Mechanics

\newpage

\begin{flushright}
{\it In Sweet Memory of Anatoly Pashnev, our collegue and friend}
\end{flushright}

\section{Introduction}

Exact solvable models have always played a distinctive role, especially in the realm of Quantum World. The use of various kinds of approximations in the course of modeling quantum phenomena is quite justified for natural materials, but it becomes a problem for designing their artificial analogues with improved or preset characteristics. Here, the exact solvability becomes a necessary theoretical tool, that allows one to predict the response of a quantum system to an external source, and, thereby, to control its behavior in different regimes of interaction with driving forces. An even more intriguing  situation is realized in the case of using, on the theoretical side, techniques for generating new exactly solvable models that have principally new properties and characteristics compared to the initial ones. Such new properties can arise as due to changes in the type of interaction between the internal ingredients of a quantum system, as well as due to the specifics of its interaction with external sources and fields.
A clear understanding, on the theory side, of the processes occurred in the system is an additional, important, though by no means the last, key to their successful implementation in practice.

Currently, it has known a sufficient number of techniques for generating new exact solutions from already known ones (see, for example, monographs and paper collections \cite{Lamb:1980ed,Eckhaus:1981tn,Drazin:1989qi,Olver:1990,Suzko:1990,Stephani:2003tm,Krasinski:1997yxj}). One of these techniques, developed since the late 40s in quantum scattering theory, is the inverse scattering method \cite{Bargmann:1949,GelLev51,Krein:1954,Marchenko:1955,Faddeev:1963}, which leads to a remarkable and non-standard result: restoring the potential from the known spectrum is an ambiguous task. What was treated as a bug before became a very useful feature after, when it were established new relations between exactly solvable potentials of Quantum Mechanics, new interconnections between non-linear equations and their solutions, symmetry structures behind and many more. After seminal papers on solitons and non-linear dynamics \cite{Zabusky:1965zz,Gardner:1967wc,Ablowitz:1974ry}, the inverse scattering method was renamed into the Inverse Scattering Transform, so that we will follow this nomenclature, irrespectively of the presence or absence of non-linearity in the system.

Another profound method of generating new solutions with exactly solvable dynamics is Supersymmetric Quantum Mechanics, whose development, after the foundation in early 80s \cite{Witten:1981nf,Cooper:1982dm,deCrombrugghe:1982nmd,Bender:1983xi,Gozzi:1977uxa}, revealed  similarities to the inverse scattering method. In particular, (almost) the same spectrum may correspond to potentials that are completely different in shape. This observation made it possible to put studies of quantum mechanical problems on a more solid ground than before, and to establish exact solutions even in the case when conventional approximations (WKB and others) failed. (See, e.g., Refs. \cite{Comtet:1985rb,Kumar:1986zy} in this respect.) The latter becomes extremely important in engineering based on Quantum Optics communication networks \cite{Hasegawa:2003, Mollenauer:2006}, as well as in designing quantum field-effect diodes and transistors \cite{Datta:2005,Tsurumi:2009}. In both cases, the employment of the Supersymmetric Quantum Mechanics \cite{Ward:2009,Berezovoj:2011ng,Berezovoj:2012ad,Macho:2018sqm,Berezovoj:2020uwm,Yim:2022} demonstrates principle advantages of the approach.

The purpose of this paper is to fill some of existed gaps in the correspondence between Supersymmetric Quantum Mechanics (SQM) \cite{Witten:1981nf,Cooper:1982dm,deCrombrugghe:1982nmd,Bender:1983xi,Gozzi:1977uxa} and the Inverse Scattering Transform (IST) \cite{Bargmann:1949,GelLev51,Krein:1954,Marchenko:1955,Faddeev:1963,Moses:1977,Defacio:1980gv,Deift:1978,Deift:1979,Abraham:1980}.  Both methods were linked \cite{Nieto:1984zj,Sukumar:1985zcg,Kwong:1985ti,Pursey:1986nc,Luban:1986iv,Berezovoi:1988tq,Sukumar:1988ch,Khare:1988is,Berezovoi:1991xc} soon after the SQM foundation and its early development,  that formed the ground for that's what we now call the SQM/IST Correspondence. The Correspondence has been well established for the stationary SQM and IST, though some points of the non-stationary extension of the SQM/IST or, more exotically, of the non-stationary extension just on one of its sides, have still remained unexplored. Here we try to take a fresh look at old problems, and to open a new avenue in studying exactly-solvable models.

The paper is structured as follows. To make the presentation self-consistent, in two subsequent sections we briefly review main points of the SQM/IST Correspondence in the stationary \cite{Nieto:1984zj,Sukumar:1985zcg,Kwong:1985ti,Pursey:1986nc,Luban:1986iv,Berezovoi:1988tq,Sukumar:1988ch,Khare:1988is,Berezovoi:1991xc} and non-stationary cases \cite{Bagrov:1990,Bagrov:1991,Bagrov:1995,Bagrov:1996jpa,Samsonov:1996,Cannata:1998ix,Schulze-Halberg:2009esa,Zelaya:2017jrf,Rasinskaite:2020vmt,Strange:2021mgx}. Formally, the material contained in this part of the paper is not new, except for the complement to the Goursat problem for the non-stationary IST, that has not been done before. Nonetheless, Sections 2 and 3 are important for understanding the main subject of our studies, and they may be considered on its own as a concise review on main ingredients of the SQM/IST Correspondence, disparate parts of which are collected in one place.

Next, in Section 4, containing the main result of the paper, we formulate rationales behind our proposal to modify the picture on the IST side. This modification, inter alia, makes it possible to pair stationary and non-stationary Hamiltonians, and to bridge the gap between ${\cal CPT}$ and ${\cal PT}$ invariant \cite{Baye:1996ytz,Bender:1998ke,Bender:1998gh,Cannata:1998bp,Dorey:2001uw,Dorey:2001hi,Znojil:2000fr,Bender:2002vv,Ahmed:2005zz,Cannata:2006htc,Rosas-Ortiz:2015zya,Cen:2018llv,Bender:2019cwm,Frith:2019nju,Bender:2023cem}
models. We prove a {\it Lemma}, according to which one may formulate the Goursat problem for the kernel of non-local transformations of wave functions, including the expression for the new paired potential as well. The proposed here exact form of the kernel for the non-stationary IST suggests a ``covariantization'' of the time derivative in the Schrodinger equation for the new paired Hamiltonian by a compensator field. Such a ``covariantization'' has been used in problems with time-dependent boundary conditions, when the wave function acquires a time-dependent geometric (i.e. the Berry-type) phase \cite{Pronin:1991}. The compensator field is introduced to balance additional contribution coming from the action of temporal derivative on the time-dependent phase. We compare our construction with that of the non-stationary SQM/IST Correspondence, and establish the relation of the proposed new kernel of the IST and its temporal derivative to the constituents forming the SQM supercharge.

Discussion of the obtained results, possible directions of their application and  further development is given in the last section. In the added to the main text Appendix, we illustrate our proposal on a classical example of a simple harmonic oscillator.

\section{The 1D SQM/IST Correspondence: Stationary case}

We begin with a review of the correspondence between the standard Supersymmetric Quantum Mechanics (SQM) \cite{Witten:1981nf,Cooper:1982dm,Bender:1983xi,Gozzi:1977uxa} and the approach of the Inverse Scattering Transform (IST) \cite{GelLev51,Krein:1954,Marchenko:1955,Faddeev:1963,Moses:1977,Defacio:1980gv,Deift:1978,Deift:1979}, developed by Abraham and Moses in \cite{Abraham:1980}. (See also Ref. \cite{Kabanikhin:2023} for a review, and Refs. therein.) Here we closely follow \cite{Nieto:1984zj}, where the correspondence between the SQM and the IST was realized. 

Specifically, the factorization of a SQM Hamiltonian requires of knowing a particular solution $\pa_x W(x)$ to the Riccati equation
\be
\left(\fr12 \pa_x W(x) \right)^2\mp \fr12 \pa^2_x W(x)=V_\pm (x),
\la{RicW}
\ee 
where $V_\pm(x)$ are the potentials for (almost) isospectral Hamiltonians $H_\pm=-\pa_x^2+V_\pm(x)$. If we fix the initial Hamiltonian as $H_+$, when the stationary Schrodinger equation $H_+ \Psi_0=E_0 \Psi_0$ is solved for
\be
\Psi_0(x)=N_0\, e^{-\fr12 W(x)},\qquad E_0=0.
\la{Psi0def}
\ee
Such a wave function is associated with the ground state of the ``bosonic'' Hamiltonian $H_+$. This Hamiltonian admits the factorization by supercharges $Q^\dag$ and $Q$, which are first order differential operators
\be
Q^\dag=-\pa_x+\fr12 \pa_x W(x),\qquad Q=\pa_x +\fr12 \pa_x W(x),
\la{Qdef}
\ee
with (derivative of) the superpotential $W(x)$. Then, the ``bosonic'' and ``fermionic'' parts of the super-Hamiltonian are constructed as
\be
H_+=Q^\dag Q,\qquad H_-=QQ^\dag,
\la{Hsuperdef}
\ee
and the spectra of $H_\pm$ are the same, except for the ground state.

Now, let's factorize the Hamiltonian $H_-$ with other supercharges, $\tilde{Q}$ and $\tilde{Q}^\dag$,
\be
\tilde{Q}=\pa_x+f(x),\qquad \tilde{Q}^\dag=-\pa_x+f(x),\qquad H_-=\tilde{Q}\tilde{Q}^\dag.
\la{tilQdef}
\ee
Taking the definition of $V_-(x)$ from eq. \rf{RicW}, we get
\be
f^2(x)+\pa_x f(x)=\left(\fr12 \pa_x W(x) \right)^2+ \fr12 \pa^2_x W(x).
\la{Ricf}
\ee
Mielnik \cite{Mielnik:1984}, after van Kampen \cite{Kampen:1971}, marked the general solution to the Riccati equation \rf{Ricf} for $f(x)$. It comes as
\be
f(x)=\fr12 \pa_x W(x)-\fr{e^{-W(x)}}{\l+\int_x^\infty dz\,e^{-W(z)}}.
\la{fSol}
\ee    
Here there has been appeared a constant $\l$, whose role will be central in what follows.

For now, we have established the correspondence between $Q$ and $\tilde{Q}$ supercharges through the ``fermionic'' part of the super-Hamiltonian $H_-$, that is, $H_-=QQ^\dag=\tilde{Q}\tilde{Q}^\dag$. Let's construct the superpartner of $H_-$ in terms of new supercharges $\tilde{Q}$, $\tilde{Q}^\dag$:
\be
\tilde{H}_+ =\tilde{Q}^\dag \tilde{Q}=-\pa^2_x+V_++2\pa_x \f(x).
\la{H+til}
\ee
Here, we have introduced
\be
\f(x)=\fr{e^{-W(x)}}{\l+\int_x^\infty dz\,e^{-W(z)}},
\la{phidef}
\ee
which obeys
\be
\pa_x \f(x)-\f^2(x)+\f(x)\pa_x W(x)=0.
\la{phieq}
\ee
Clearly, $\tilde{H}_+\ne H_+$, and $\tilde{H}_+$ corresponds to the Hamiltonian with new potential 
\be
\tilde{V}_+=V_++2\pa_x \f(x).
\la{VtilSQMdef}
\ee 
But what about the spectra of these two Hamiltonians?

It turns out that the Hamiltonians $H_+$ and $\tilde{H}_+$ are also (almost) isospectral, via their parent Hamiltonian $H_-$. Indeed, we can write
\[
\tilde{H}_+\tilde{Q}^\dag=\left(\tilde{Q}^\dag \tilde{Q} \right)\tilde{Q}^\dag=\tilde{Q}^\dag \left(\tilde{Q}\tilde{Q}^\dag \right)=\tilde{Q}^\dag H_- ,
\] 
that means that eigenstates of $\tilde{H}_+$ are determined by $|\tilde{\y}^{(+)}_i\rangle=\tilde{Q}^\dag|\y^{(-)}_i\rangle$, where $|\y^{(-)}_i\rangle$ are the eigenstates of $H_-$. So that, the energies (eigenvalues of $\tilde{H}_+$) will be given by $E_i$, $i=1,\dots,N$. However, as in the case of $H_+$, one needs to add the ground state $\tilde{Q}|\tilde{\y}^{(+)}_0\rangle=0$ \cite{Mielnik:1984}. In the coordinate representation, the corresponding wave function $\tilde{\Psi}_0$ looks as
\be
\tilde{\Psi}_0=\tilde{N}_0 \,e^{-\fr12 W(x)} e^{\int_0^x dz\,\f(z)} .
\la{Psi0tildef}
\ee
It is straightforward to verify that $\tilde{H}_+\tilde{\Psi}_0=0$, hence the energy of the ground state of $\tilde{H}_+$ is also $\tilde{E}_0=E_0=0$.

Other eigenstates of two Hamiltonians $H_+$ and $\tilde{H}_+$ are related to each other via
\be
|\tilde{\y}^{(+)}_i\rangle=\tilde{Q}^\dag Q|\y^{(+)}_i\rangle ,
\la{tilpsi+topsi+}
\ee
so that the normalizations of wave functions $\{\Psi_0,\y^{(+)}_i\}$ and $\{\tilde{\Psi}_0,\tilde{\y}^{(+)}_i\}$ are different. This situation, when the eigenvalues of two spectra are equal, but the normalizations of a finite number of point spectra wave functions are different, is well-known in the IST, and it has been considered, e.g., in \cite{Deift:1978,Deift:1979,Abraham:1980}. 

According to Abraham\,\&\,Moses, Ref. \cite{Abraham:1980}, the wave functions of two isospectral Hamiltonians are related to each other via non-local transformation
\be
\tilde{\y}^{(+)}_i(x)=\y^{(+)}_i(x)+\int_x^\infty dy\,K(x,y)\y^{(+)}_i(y),
\la{AMtrans}
\ee
with the kernel of the transformation $K(x,y)$. Both eqs. \rf{tilpsi+topsi+} and \rf{AMtrans} are different realizations of $|\tilde{\y}^{(+)}_i\rangle=U|\y^{(+)}_i\rangle$, with local (eq. \rf{tilpsi+topsi+}) and non-local (eq. \rf{AMtrans}) transformation operators $U$, respectively.

And, as it became clear after the results of \cite{Nieto:1984zj}, the kernel of the non-local transformations \rf{AMtrans} should be taken to be
\be
K(x,y)=-\fr{\Psi_0(x)\Psi_0(y)}{\Lambda+\int_x^\infty dz |\Psi_0(z)|^2},
\la{Kxydef}
\ee 
with a constant $\Lambda$, that (see \cite{Abraham:1980} for details) makes it possible to construct a new, isospectral, Hamiltonian with the potential
\be
\tilde{V}_+=V_+-2 \pa_x K(x,x) .
\la{VtilISTdef}
\ee
Comparing two equations, eq. \rf{VtilISTdef} and eq. \rf{VtilSQMdef}, to each other, one may conclude that $K(x,x)=-\f(x)$. Hence, eq. \rf{Kxydef} turns into eq. \rf{phidef} as soon as (cf. eq. \rf{Psi0def}) $\l=\Lambda/N^2_0$. In addition, we have to check the following equation on the kernel 
\be
\pa^2_x K(x,y)-\pa^2_y K(x,y)+2\left[\pa_x K(x,x)\right] K(x,y)-\left(V_+(x)-V_+(y) \right)K(x,y)=0,
\la{Keqdef}
\ee
which holds for the kernel \rf{Kxydef}. Other aspects of the SQM/IST correspondence, including the general non-equivalence of the approaches, can be found, e.g., in \cite{Sukumar:1985zcg,Kwong:1985ti,Pursey:1986nc,Luban:1986iv,Mielnik:2004cg}. Note that for confined potentials, with discrete (point in the nomenclature of \cite{Abraham:1980}) spectra, the equivalence of the Abraham-Moses to the SQM approach was confirmed in \cite{Berezovoi:1988tq,Berezovoi:1991xc}. So that we will focus just on this case.

Thus, different approaches to the construction of new Hamiltonians -- SQM and the IST -- are inherently related. And we can use this correspondence between two approaches in both directions. Here we have started with the SQM Hamiltonian construction, and turned to the IST technique after. We can use the reverse procedure equally. That is, we can begin with the IST approach first, to make the conclusion on the structure of the paired SQM Hamiltonian after that.

\section{The 1D SQM/IST Correspondence: Non-stationary case}

Elaborating on the construction of new isospectral Hamiltonians with time-dependent potentials, we have to sum up the previous achievements in the stationary case. First, there is the correspondence between two different approaches -- the Inverse Scattering Theory and Supersymmetric Quantum Mechanics -- that are used to generate a new isospectral Hamiltonian ($\tilde{H}_+$ in the case) to the initial one ($H_+$ in the used here notation). This correspondence is exact for confined potentials that includes the harmonic potential as well. 

Second, the SQM/IST correspondence is realized quite differently on the both sides. As we have noticed before, the relation between eigenstates (wave functions) of $\tilde{H}_+$ and $H_+$ is established either by two consequent local transformations with different supercharges (cf. eq.\rf{tilpsi+topsi+}), or by the single non-local transformation with the integral kernel (cf. eq. \rf{AMtrans}). Any of these transformations include specific combinations of the ground state wave functions, eqs. \rf{phidef} or \rf{Kxydef}, with some unspecified, and generally complex-valued,  constant $\l$. (Recall, $\Lambda$ in \rf{Kxydef} is related to $\l$ in \rf{phidef} via the square of the normalization constant $N_0$.) The natural requirement that restricts this constant is 
\be
\l+\int_x^\infty dz \,e^{-W(x)}\ne 0.
\la{lambda}
\ee
For the rest, this constant can be chosen freely in the domain of the normalizability of the $\tilde{H}_+$ wave functions \cite{Sukumar:1988ch}. We will turn to the discussion of this aspect of $\l$ later on, after reviewing the non-stationary extension of the SQM/IST correspondence. 

And third, the considered in \cite{Nieto:1984zj} SQM/IST correspondence works in the stationary case. That is, we have the correspondence between isospectral Hamiltonians, entering the stationary Schrodinger equations. Below, we will extend this picture on two notable cases, when the correspondence is established as between non-stationary Hamiltonians, as well as between stationary and non-stationary Hamiltonians.

Let's begin with the SQM/IST correspondence for two non-stationary isospectral Hamiltonians. The time-dependent extension on the IST side and its relation to the (S)QM approach has been elaborated in \cite{Bagrov:1990,Bagrov:1991}. There, it was proposed the following extension of the kernel $K(x,y)$:
\be
K(x,y;t)=-\fr{\y^{(+)}(x,t)(\y^{(+)}(y,t))^*}{\l+\int^\infty_x dz\,|\y^{(+)}(z,t)|^2}
\la{Kxytdef}
\ee 
with constant $\l$. Using the developed in \cite{Abraham:1980} technique (see Appendix of \cite{Abraham:1980} for details on the stationary case), one can set up the Goursat problem as
\be
\left\{
\begin{array}{c}
\tilde{V}_+(x,t)=V_+(x,t)-2\pa_x K(x,x;t),\\
\\
\pa^2_x K(x,y;t)-\pa^2_y K(x,y;t)+i\pa_t K(x,y;t)-(\tilde{V}_+(x,t)-V_+(y,t))K(x,y;t)=0 .
\end{array}
\right.
\la{GoursatT}
\ee    
It is straightforward to verify that the kernel \rf{Kxytdef} satisfies the second equation of \rf{GoursatT} for wave functions of the complete Schrodinger equation with $H_+$,
\be
\left(H_+-i\pa_t \right)\y^{(+)}(x,t)=0,\qquad H_+=-\pa^2_x+V_+(x,t).
\la{SchroNS}
\ee
The new potential $\tilde{V}_+(x,t)$ enters the twinned with $H_+$ Hamiltonian $\tilde{H}_+=-\pa^2_x+\tilde{V}(x,t)$. Its wave functions $\tilde{\y}^{(+)}(x,t)$ also obey the non-stationary Schrodinger equation, now with $\tilde{H}_+$:
\be
\left(\tilde{H}_+-i\pa_t \right)\tilde{\y}^{(+)}(x,t)=0,\qquad \tilde{H}_+=-\pa^2_x+\tilde{V}_+(x,t).
\la{SchroNStil}
\ee

The SQM side of the story was developed in Refs. \cite{Bagrov:1990,Bagrov:1991,Bagrov:1995,Bagrov:1996jpa,Samsonov:1996}. (See also Refs. \cite{Cannata:1998ix,Schulze-Halberg:2009esa,Suzko:2009zz,Zelaya:2017jrf,Frith:2019nju,Rasinskaite:2020vmt,Strange:2021mgx} for the development and applications of methods by Bagrov et al.) The supercharge $Q$, associated with $H_+$, is formed from the superpotential $W(x,t)$, similarly to eqs. \rf{Qdef}. Concerning the supercharge $\tilde{Q}$, it is realized in the following way (see Refs. \cite{Schulze-Halberg:2009esa,Suzko:2009zz} for the detailed analysis of the construction of \cite{Bagrov:1990,Bagrov:1991,Bagrov:1995,Bagrov:1996jpa,Samsonov:1996}):
\be
\tilde{Q}=l(t)\left(\pa_x+f(x,t) \right),\qquad f(x,t)=\fr12 \pa_x W(x,t)-\f(x,t).
\la{tilQt}
\ee
The functions $l(t)$ and $\f(x,t)$ are generally complex-valued functions. Then, the intertwining relation, which is a paraphrase of the isospectrality,
\be
\left(i\pa_t -H_- \right)\tilde{Q}=\tilde{Q}\left(i\pa_t-\tilde{H}_+\right)
\la{intertwin}
\ee
makes it possible to derive the relation \cite{Bagrov:1990,Bagrov:1991} between the new and the old potentials,
\be
\tilde{V}_+(x,t)=V_+(x,t)+2\pa_x \f(x,t)-i\pa_t \ln l(t).  
\la{VtilCSE}
\ee
And the function $\f(x,t)$ is restricted to satisfy 
\be
\pa_x \left[f^2(x,t)-\pa_x f(x,t)-\tilde{V}_+(x,t) \right]-i\pa_t f(x,t)=0.
\la{feqCSE}
\ee

Note also that the requirement of real-valued potentials in \rf{VtilCSE} leads to the following constraints:
\be
\pa_t \ln |l(t)|^2=-4 \pa_x \left[ \Im\text{m}\, \f(x,t)\right] \qquad \leadsto \qquad \pa^2_x \left[ \Im\text{m}\, \f(x,t)\right]=0,
\la{constraints}
\ee
so that the non-triviality of $l(t)$ requires a complex-valued $\f(x,t)$. At the same time, dealing with a real-valued potential to get real-valued eigenvalues of the Hamiltonian is an excess requirement \cite{Baye:1996ytz,Bender:1998ke,Bender:1998gh}. We will turn to the more detailed discussion of this point in the next section. 

For now let us note that, from the construction of the kernel and supercharges, it is clear that we are dealing with non-stationary paired Hamiltonians. The SQM/IST correspondence at the level of the supercharge/kernel relation is established by comparing the right hand sides of $\tilde{V}_+(x,t)$ in \rf{GoursatT} and \rf{VtilCSE}. As a result, we obtain a relationship between the spatial derivative of the transformation kernel and the derivatives, both spatial and temporal, of the basic ingredients of the supercharge. Thus, in contrast to the stationary case, the SQM/IST correspondence is realized indirectly, through the derivatives of the main quantities.

However, the direct realization of the SQM/IST correspondence is restored within a slightly different, but important, extension of the stationary case, which will formally connect stationary and non-stationary Hamiltonians. Now, we are turning to the assembling of this construction.

\section{The 1D SQM/IST Correspondence: Stationary to non-stationary Hamiltonians pairing}

Let's carry on with a brief discussion on the kernel structure for the stationary case. Equation \rf{Kxydef} can be generalized to
\be
K(x,y)=-\fr{C_2 \Psi_0(x)\Psi_0(y)}{C_1+C_2 \int^\infty_x dz\,|\Psi_0(z)|^2},\qquad C_{1,2}=\text{const} \in \mathbb{C}.
\la{KxyC12}
\ee
However, what matters is the ratio of these two constants, which we have denoted as $\l=C_1/C_2$. {\it Changing the $\l$ does not affect the spectrum}, affecting the potential and wave functions, respectively. Therefore, we can instantly change the $\l$ in its value, within its range, determined by relation \rf{lambda}. 
Or, we can relate any instant $\l$ changing to some unique value of the time parameter. Smooth changes form a function of time $\l(t)$ after that. Thus, we would like to investigate consequences of using the following kernel template on the IST side:
\be
K(x,y;t)=-\fr{C_2 \Psi_0(x)\Psi_0(y)}{C_1(t)+C_2 \int^\infty_x dz\,|\Psi_0(z)|^2},\quad C_2=\text{const},
\la{KxyC12t}
\ee
where $\Psi_0(x)$ is the ground state wave function of $H_+$ (cf. eq. \rf{Psi0def}). 

With the kernel of non-local (integral) transformations of wave functions \rf{KxyC12t}, we expect that the supercharges from eqs. \rf{tilQdef} will become functions of time, via  
\be
\f(x,t)=\fr{e^{-W(x)}}{\l(t)+\int^\infty_x dz\,e^{-W(z)}}.
\la{phitdef}
\ee
Then, the time dependence of the $\tilde{H}_+$ ground state wave function,
\be
\tilde{\Psi}_0(x,t)=\tilde{N}_0 e^{-\fr12 W(x)}e^{\int^x_0 dz\,\f(z,t)},
\la{Psi0tilt}
\ee
can be treated as a time-dependent geometric phase, i.e.,
\be
\tilde{\Psi}_0(x,t)=\tilde{N}_0 e^{-\fr12 W(x)}e^{i\Phi(x,t)},\quad 
i\Phi(x,t)=\int^x_0 dz\,\f(z,t).
\la{Psi0tilt1}
\ee
Recall, wave functions with time-dependent geometric phase naturally appear in quantum systems with time-dependent boundary conditions. And a by-product of the consideration is the ``covariantization'' \cite{Pronin:1991} of the non-stationary Schrodinger equation in temporal direction.

Taking the latter into account, let us formulate the considered problem in more general terms. 

\bsk
{\it Lemma.} Suppose that two sets of wave functions, $\y_\ve(x)$ and $\Psi(x,t)$, are related to each other via non-local transformations
\be
\Psi(x,t)=e^{-i\ve t}\y_\ve(x)+\int_x^\infty dy\,K(x,y;t)\, e^{-i\ve t}\y_\ve(y).
\la{PsiNSn}
\ee  
The kernel of these transformations  
\be
K(x,y;t)=-\fr{C_2 \y(x,t) \y^*(y,t)}{C_1(t)+C_2\int^\infty_x dz |\y(z,t)|^2} ,
\la{Kxytalt1}
\ee
is constructed out of the stationary wave functions $\y(x,t)=e^{-i\ve t} \y_\ve(x)$, which obey the Schrodinger equation
\be
\left(H_0 -i\pa_t \right) \y(x,t)=0, \qquad H_0=-\pa^2_x+V(x).
\la{NSE}
\ee
If the wave function $\Psi(x,t)$ obeys the generalization of the Schrodinger equation,
\be
\left( H-iD_t \right) \Psi(x,t)=0,\qquad H=-\pa^2_x+U(x,t) ,
\la{NSEn}
\ee
in which $D_t=\pa_t+A_t $ is a ``covariantization'' of the time derivative by a ``compensator'' field $A_t$, determined by $D_t K(x,y;t)=0$, then the relation between the paired potentials and the Master equation for the kernel (the Goursat problem) looks as follows:
\be
\left\{
\begin{array}{c}
U(x,t)=V(x)-2\pa_x K(x,x;t)+iA_t,\\
\\
\pa^2_x K(x,y;t)-\pa^2_y K(x,y;t)+iD_t K(x,y;t)-\left(\left[U(x,t)-iA_t\right]-V(y)\right)K(x,y;t)=0 .
\end{array}
\right.
\la{NSEGoursatalt}
\ee

\bsk
{\it Proof.} Let's act with $(H-iD_t)$ operator on the wave function \rf{PsiNSn}. This operation can be split in two parts. One of them is
\[
\left(H-iD_t \right)\y(x,t)=\left(-\pa^2_x+U(x,t)-i\left(\pa_t+A_t \right) \right)e^{-i\ve t}\y_\ve(x)
\]
\be
=\left(U(x,t)-iA_t-V(x) \right) e^{-i\ve t}\y_\ve(x) ,
\la{0n}
\ee
where we have used eq. \rf{NSE}. The other part of eq. \rf{NSEn} contains
\[
\left(H-iD_t \right)\int_x^\infty dy\, K(x,y;t) e^{-i\ve t} \y_\ve(y)=\left(-\pa^2_x+U(x,t)-iD_t \right)\int_x^\infty dy\, K(x,y;t) e^{-i\ve t} \y_\ve(y)
\]
\[
=\left(-\pa^2_x+U(x,t)\right)\int_x^\infty dy\, K(x,y;t) e^{-i\ve t} \y_\ve(y)-i\int^\infty_x dy\, \underbrace{\left(D_t K(x,y;t) \right)}_{=0}e^{-i\ve t} \y_\ve(y)
\]
\[
-i \int^\infty_x dy\, K(x,y;t) \left(\pa_t+A_t\right)e^{-i\ve t} \y_\ve(y)
\]
\[
=\left(-\pa^2_x+U(x,t)-\ve \right)\int_x^\infty dy\, K(x,y;t) e^{-i\ve t} \y_\ve(y)-i\int^\infty_x K(x,y;t) A_t e^{-i\ve t} \y_\ve(y).
\]
By use of 
\[
\left(-\pa^2_x+U(x,t)-\ve \right)\left(\y_\ve(x)+ \int^\infty_x dy\, K(x,y) \y_\ve(y)\right)=\left(U(x,t)-V(x) \right) \y_\ve(x)+\y_\ve(x) \fr{d}{dx}K(x,x)
\]
\[
+\y_\ve(x)\underbrace{\left( \fr{\pa}{\pa x}K(x,y)\Big|_{y=x}+\fr{\pa}{\pa y} K(x,y)\Big|_{x=y}\right)}_{=\fr{d}{dx}K(x,x) }
\]
\be
-\int^\infty_x dy\, \left(\pa^2_x K(x,y)-\pa^2_y K(x,y)-U(x,t)K(x,y)+V(y)K(x,y) \right)\y_\ve(y), 
\la{PsiSE2}
\ee
we finally arrive at
\[
\left(H-iD_t \right)\Psi(x,t)=e^{-i\ve t}\Bigg[\left(U(x,t)-iA_t-V(x)+2\fr{d}{dx}K(x,x;t) \right) \y_\ve(x)
\]
\[
-\int^\infty_x dy\, \left(\pa^2_x K(x,y;t)-\pa^2_y K(x,y;t)+iA_t K(x,y;t)-\left(U(x,t)-V(y)\right)K(x,y;t) \right)\y_\ve(y,t)\Bigg].
\]
Equating to zero the r.h.s. of the latter equation, we get the system of equations \rf{NSEGoursatalt}. \,\,\,$\blacksquare$

\bsk
In the notation of the paper, we have proved that the usage of the kernel \rf{KxyC12t} results in the construction of the new, paired to $H_+$, effective Hamiltonian $\tilde{H}_+$, extended by the interacting with the ``compensator'' $A_t$ term. This field is fixed by the requirement of having the ``covariantly constant'' kernel, i.e.,
\be
\pa_t K(x,y;t)+A_t K(x,y;t)=0.
\la{covDK}
\ee
Then, for $K(x,y;t)$ \rf{Kxytalt1}, 
\be
\pa_t K(x,y;t)=-\fr{\pa_t C_1(t)}{C_1(t)+C_2\int^\infty_x dz\, \y_\ve(z)^2} K(x,y;t)=- A_t K(x,y;t)
\la{DtKsol}
\ee
so that the relation between the initial and the final, after the IST, potential in our case (with the kernel \rf{KxyC12t}) becomes
\be
\tilde{V}_+(x,t)=V_+(x)-2\pa_x K(x,x;t)+\fr{i\pa_t C_1(t)}{C_1(t)+C_2\int^\infty_x dz\, |\Psi_0(z)|^2}\,.
\la{tilVS2NS}
\ee 
Comparing the latter expression with that of eq. \rf{VtilCSE}, we can establish the direct correspondence between the IST and the SQM quantities:
\be
K(x,x;t)=-\f(x,t),\qquad  \pa_t \ln l(t)=-\fr{\pa_t C_1(t)}{C_1(t)+C_2\int^\infty_x dz\, |\Psi_0(z)|^2}.
\la{Kphil}
\ee

Therefore, the usage of the special IST kernel \rf{KxyC12t} makes it possible: 1) to construct a new paired time-dependent potential from the initially stationary one; 2) to realize the SQM/IST correspondence directly in the non-stationary case, with the special kernel of the IST \rf{Kxytalt1}. 

Another interesting feature behind our proposal can be figured out from the analysis of the potential \rf{tilVS2NS}. The requirement to have a real-valued potential $\tilde{V}(x,t)$ leads to the following relation between the kernel and the ``compensator'' field:
\be
\Re\text{e} A_t(x,t)=2\pa_x \left(\Im\text{m}\, K(x,x;t) \right).
\la{AKrel}
\ee
Hence, the kernel $K(x,y;t)$ \rf{Kxytalt1} has to be complex-valued, to provide the non-triviality of $A_t(x,t)$. For the IST from a stationary system to a non-stationary one, it means that the parameter $\l(t)=C_1(t)/C_2$ has to be a complex-valued function of time. Then, according to \rf{PsiNSn} and \rf{Kxytalt1}, wave functions of the paired Hamiltonian are principally non-stationary ones, in the sense that they are non-factorizable in time and space coordinates, and they are complex functions. If so, they correspond to scattering states, so that the IST \rf{PsiNSn} with the kernel \rf{Kxytalt1} may, for instance, transform bound states to scattering states. 

Even more interesting situation takes place when we omit the restriction on the new scattering potential to be a real-valued function of time and space coordinates. 
In this case we still have a possibility to operate with Hamiltonians having real-valued energy spectra. It happens, for example, for the so-called ${\cal P}{\cal T}$-invariant Hamiltonians \cite{Bender:1998ke,Bender:1998gh,Cannata:1998bp,Dorey:2001uw,Dorey:2001hi,Znojil:2000fr,Bender:2002vv,Ahmed:2005zz,Cannata:2006htc,Rosas-Ortiz:2015zya,Cen:2018llv}, properties of which are nicely reviewed in Refs. \cite{Bender:2019cwm}, \cite{Frith:2019nju} and \cite{Bender:2023cem}.

Let's apply the ${\cal P}{\cal T}$ transformations to the potential \rf{tilVS2NS}. Under the ${\cal P}{\cal T}$ transformations, $V_+(x) \ra V_+(-x)$. For $K(x,y;t)$ we have (recall, $C_2$ is a constant)
\[
K(x,y;t)\ra -\fr{C_2 \y_\ve(-x)\y_\ve(-y)}{C_1(-t)+C_2\int_{-x}^{-\infty} d(-z)\, \y_\ve(-z)^2}=- \fr{C_2 \y_\ve(x)\y_\ve(y)}{C_1(-t)-C_2\int_{x}^{\infty} dz\, \y_\ve(z)^2}  
\]
\[
=- \fr{C_2 \y_\ve(x)\y_\ve(y)}{C_1(-t)-C_2\int_{x}^{\infty} dz\, \y_\ve(z)^2}\,. 
\]
Hence, ${\cal P}{\cal T}\left[K(x,y;t)\right] \ra -K(x,y;t)$ for $C_1(-t)=-C_1(t)$, that means ${\cal P}{\cal T}\left[\pa_x K(x,y;t)\right] \ra \pa_x K(x,y;t)$.

Upon the action of ${\cal P}{\cal T}$ operators on the imaginary part of a complex-valued potential \rf{tilVS2NS} we obtain
\be
{\cal P}{\cal T} \left[\Im\text{m}\, \tilde{V}_+(x,t)\right]=-\fr{\pa_t C_1(-t)}{C_1(-t)-C_2\int^\infty_x dz\, \y_\ve(z)^2}\,.
\la{ImU}
\ee 
And if $C_1(-t)=-C_1(t)$, the $i\Im\text{m}\,\tilde{V}_+(x,t)$ stays to be ${\cal P}{\cal T}$ invariant.

Therefore, the restriction on $C_1(t)$ to be an odd function of time provides, together with the even parity of the potential $V_+(x)$, the invariance of the paired potential under the ${\cal P}{\cal T}$ transformations. And the ${\cal P}{\cal T}$ invariance is not a drawback for the factorization and application of Supersymmetry to the consideration. (Cf., e.g., Refs. \cite{Dorey:2001hi,Znojil:2000fr,Cannata:2006htc,Rosas-Ortiz:2015zya,Cen:2018llv,Frith:2019nju} in this respect.) Clearly, the supercharge of the paired Hamiltonian, $\tilde{Q}=l(t)\left(Q-\f(x,t)\right)$, is constructed out of the initial even-parity Hamiltonian supercharge $Q$, and the determined by eqs. \rf{Kphil} functions with $C_1(-t)=-C_1(t)$. (See Appendix for details on the paired to a simple harmonic oscillator ${\cal PT}$-invariant Hamiltonian.) Hence, in the case of real-valued energy Hamiltonians, we have the SQM/IST Correspondence between
\begin{enumerate}[label=\roman*)]
\item
stationary and non-stationary Hamiltonians;
\item
bound, scattering and tunnelling states;
\item
${\cal CPT}$-invariant and ${\cal PT}$-invariant Hamiltonians.
\end{enumerate}
Early, relations between ${\cal CPT/PT}$-invariant Hamiltonians were established merely for non-stationary systems. (See, e.g., \cite{Cen:2018llv,Frith:2019nju}, and Refs. therein.)  

Furthermore, even if all the restrictions on the potential (like the requirement of dealing with real-valued or ${\cal PT}$-invariant potentials) will be omitted, we will take another possibility to construct paired complex-valued Hamiltonians. As it is known from classical electrodynamics, choosing specific boundary conditions leads to the appearance of quasi-normal modes with complex-valued frequencies. The complex-valued frequencies correspond to complex energies even for a real-valued potential. Therefore, the pairing between real-valued (stationary) and complex-valued (non-stationary) potentials may be realized as a result of imposing the specific boundary conditions. The quasi-normal modes play, in particular, an important role in waveguides, resonators, and in Black Hole Physics as well. So that, the proposed here mechanics of generation of new exactly-solvable potentials may also find its natural application in these domains of study.

\section{Discussion and final remarks}

In summary, we propose a new mechanism for generating paired potentials based on a very specific time-dependent extension of the IST kernel or the SQM supercharge. The standard generalization of the Abraham-Moses construction \cite{Abraham:1980} to the non-stationary case \cite{Bagrov:1990,Bagrov:1991,Bagrov:1995} assumes the replacement of stationary wave-functions with non-stationary ones. At the same time, the integration constant, coming from the general solution to the Riccati equation \cite{Mielnik:1984,Kampen:1971} and entering either the IST kernel or the SQM supercharge, remains unchanged.  Our proposal is based on the opposite consideration, when the integration constant becomes a function of time, while the rest of the IST kernel or the SQM supercharge remains the same as in the stationary case. So that, if the standard non-stationary generalization of the Abraham-Moses construction pairs time-dependent Hamiltonians, our approach admits a non-standard and non-trivial pairing of time-independent and time-dependent Hamiltonians.

Other non-trivial features of our proposal consist in the correspondence between real-valued and complex-valued potentials, since the elaborated here procedure of the paired Hamiltonian construction results, generally, in the appearance of a complex-valued potential after the IST employment. If the complex-valued potential is not constrained, the eigenvalues of a complex-valued Hamiltonian are usually complex. And it can be viewed in different ways. For instance, the complex part of energy values may indicate the dissipation in real physical systems, that describes the natural energy loss, e.g., in Quantum Optics, or in optical fibers. Another way to deal with complex-valued energies is to consider effects of tunnelling through barriers suggested by the potential shape, or by specific boundary conditions. Note that the role of boundary conditions in the twinning of Hamiltonians with real-valued and complex-valued potentials becomes fundamental, since setting the specific boundary conditions is a most natural way to generate quasi-normal modes with complex-valued frequencies. Therefore, following the proposed here recipe, one may construct, on the ground of a hermitian stationary exactly-solvable Hamiltonian, the effective exactly-solvable time-dependent Hamiltonian for a system with the energy dissipation.

The requirement to have real-valued energies in the spectrum of a complex-valued Hamiltonian leads us to the ${\cal PT}$-invariant quantum mechanics. (See, e.g., \cite{Bender:2019cwm,Frith:2019nju, Bender:2023cem} and Refs. therein.) We established the set of constraints for the initial hermitian time-independent Hamiltonian potential and the kernel of the non-local IST, that leads to the time-dependent complex-valued paired potential. The standard SQM and, partially, the ${\cal PT}$-invariant SQM gained in popularity in Solitons Theory \cite{Koller:2005,Kevrekidis:2015}, Optical Waveguides \cite{Ward:2009,Macho:2018sqm}, models of Analogue Black Holes \cite{Bagchi:2023vzx},  diffusion theory \cite{Berezovoj:2010zz,PeglowBorges:2010zz} and many more. (See, for instance, Ref. \cite{FernandezC:2018cdo} for a review on modern trends in SQM.) So we hope that the obtained here outcomes of our work will give an impact on investigations in the mentioned as well as in other branches of modern science and technology.

Finally, let's pay special attention to the application of our results in Black Hole Physics. We have mentioned the Quasi-Normal Modes, which, together with the phenomenon called Superradiation, play a very important role in modern studies of Black Holes. A time ago it was established the isospectrality of the Regge-Wheeler and Zerilli equations for spin-2 perturbations over a specified gravitational background \cite{Glampedakis:2017rar}. However recently, in \cite{Li:2023ulk}, it was observed that this phenomenon does not properly work for a general gravitational background. It would be interesting to investigate, on the ground of the obtained by us results, the (non-)isospectrality between equations for spin-2 perturbations over stationary and non-stationary gravitational background with formally complex-valued effective potentials. We hope to report on this and other outcomes of our work in future communications.

\bsk
{\bf Acknowledgements}. Different aspects related to the subject of the paper and much more were once discussed with Anatoly Pashnev. We had the opportunity and the privilege to know him and to work with him. And we never forget his kindness and supportive spirit. His unexpected and premature passing was a great lost for us. We dedicate this article to his sweet memory. 

The work of A.J.N. is supported in part within the Cambridge-NRFU 2022 initiative ''Individual research (developments) grants for researchers in Ukraine (supported by the University of Cambridge, UK)'', project №2022.02/0052.

\bsk\bsk

\begin{appendix}
\numberwithin{equation}{section}

\section{An example of pairing of ${\cal CPT/PT}$-symmetric Hamiltonians}

Consider, for instance, the Hamiltonian of a simple harmonic oscillator:
\be
H_+=-\pa^2_x+x^2.
\la{HOH}
\ee 
The stationary Schrodinger equation $H_+ \y_n=\e_n \y_n$ (with physical boundary conditions) is solved with
\be
\y_n(x)=\fr1{2^{n/2}(n!)^{1/2}\pi^{1/4}}\,e^{-x^2/2}H_n(x),\qquad n=0,1,2,\dots
\la{psiHO}
\ee
$H_n(x)$ is the Hermit polynomial of $n$-th order; the eigenvalues of \rf{HOH} are $\e_n=2n+1$.

The IST kernel is constructed out the wave functions \rf{psiHO} according to \rf{KxyC12t}:
\be
K(x,y;t)=-\fr{1}{\sqrt{\pi}} \fr{e^{-\fr12 (x^2+y^2)}}{\lambda(t)+\fr12 \text{Erfc(x)}},
\la{KxytHO}
\ee
where the $\text{Erfc(x)}$ function is defined by
\be
\text{Erfc(x)}=\fr{2}{\sqrt{\pi}}\,\int_x^{\infty} dz\, e^{-z^2}.
\la{Erfcdef}
\ee
Therefore (cf. \rf{DtKsol}),
\[
\pa_t K(x,y;t)=-\fr{\pa_t \lambda(t)}{\lambda(t)+\fr12 \text{Erfc(x)}}\,K(x,y;t)=-A_t  K(x,y;t) \quad \leadsto
\]
\be
A_t=\fr{\pa_t \lambda(t)}{\lambda(t)+\fr12 \text{Erfc(x)}} .
\la{AtHO}
\ee

By taking eqs. \rf{AtHO} and \rf{tilVS2NS} into account, the paired potential $\tilde{V}_+$ looks as follows:
\be
\tilde{V}_+(x,t)=x^2+2\left[\left(\fr{e^{-x^2}}{\sqrt{\pi} \left(\lambda(t)+\fr12 \text{Erfc}(x)\right)}-1\right)^2-1\right]+\fr{i\pa_t \lambda(t)}{\lambda(t)+\fr12 \text{Erfc(x)}}\,.
\la{tilVHO}
\ee
Choosing $\lambda(t)=\sin(t)$ to get a ${\cal PT}$-invariant potential with real eigenvalues for the corresponding Hamiltonian, we finally arrive at
\be
\tilde{V}_+(x,t)=x^2+2\left[\left(\fr{e^{-x^2}}{\sqrt{\pi} \left(\sin(t)+\fr12 \text{Erfc}(x)\right)}-1\right)^2-1\right]+\fr{i\cos(t)}{\sin(t)+\fr12 \text{Erfc(x)}}\,.
\la{tilVHO1}
\ee

The supercharge $\tilde{Q}$ is constructed according to eqs. \rf{tilQt} and \rf{Kphil}. In the considered case it becomes
\be
\tilde{Q}=2 e^{-x^2}-2\left(\sin t+\fr12 \text{Erfc}(x) \right)\left(x+\pa_x \right).
\la{QtilHO}
\ee
For $\tilde{Q}^\dag$ we get, respectively,
\be
\tilde{Q}^\dag=2 e^{-x^2}-2\left(\sin t+\fr12 \text{Erfc}(x) \right)\left(x-\pa_x \right),
\la{QtildagHO}
\ee
so that the ${\cal PT}$-invariant paired Hamiltonian is $\tilde{H}_+=\tilde{Q}^\dag \tilde{Q}+iA_t$.

\end{appendix}

\end{document}